\documentclass[acmlarge]{acmart}
\AtBeginDocument{%
  }

\setcopyright{acmlicensed}
\copyrightyear{2026}
\acmYear{2026}
\acmDOI{XXXXXXX.XXXXXXX}

\acmJournal{JOCCH}
\acmVolume{1}
\acmNumber{1}
\acmArticle{1}
\acmMonth{1}

\usepackage{graphicx}
\usepackage[inline]{enumitem}
\usepackage{hyperref}

\begin{document}

\title[Design Space of RAG-Based Avatars for Archaeology]{Design Space and Implementation of RAG-Based Avatars for Virtual Archaeology}

\author{Wilhelm Kerle-Malcharek}
\authornote{Both authors contributed equally to this research.}

\email{wilhelm.kerle@uni-konstanz.de}
\orcid{0009-0001-3415-1136}
\affiliation{%
  \institution{Dept.~of Computer and Information Science, University of Konstanz}
  \city{Konstanz}
  \country{Germany}
}

\author{Giulio Biondi}
\orcid{0000-0002-1854-2196}
\authornotemark[1]
\email{giulio.biondi@unipg.it}
\affiliation{%
  \institution{CeDiPa, University of Perugia}
  \city{Perugia}
  \country{Italy}
}

\author{Karsten Klein}
\email{karsten.klein@uni-konstanz.de}
\orcid{0000-0002-8345-5806}
\affiliation{%
  \institution{Dept.~of Computer and Information Science, University of Konstanz}
  \city{Konstanz}
  \country{Germany}
}

\author{Ulf Hailer}
\orcid{0000-0002-1391-0703}
\email{ulf.hailer@uni-konstanz.de}
\affiliation{%
  \institution{Dept.~of History, University of Konstanz}
  \city{Konstanz}
  \country{Germany}
}

\author{Steffen Diefenbach}
\orcid{0009-0008-6594-2329}
\email{steffen.diefenbach@uni-konstanz.de}
\affiliation{%
  \institution{Dept.~of History, University of Konstanz}
  \city{Konstanz}
  \country{Germany}
}

\author{Fabrizio Grosso}
\orcid{0000-0002-5766-4567}
\email{fabrizio.grosso@unipg.it}
\affiliation{%
  \institution{CeDiPa, University of Perugia}
  \city{Perugia}
  \country{Italy}
}

\author{Marco Legittimo}
\orcid{0000-0002-4532-7678}
\email{marco.legittimo@unipg.it}
\affiliation{%
  \institution{CeDiPa, University of Perugia}
  \city{Perugia}
  \country{Italy}
}

\author{Paola Venuti}
\orcid{0009-0007-5006-9590}
\email{paola.venuti@unipg.it}
\affiliation{%
  \institution{CeDiPa, University of Perugia}
  \city{Perugia}
  \country{Italy}
}

\author{Carla Binucci}
\email{carla.binucci@unipg.it}
\affiliation{%
  \institution{Dept.~of Engineering, University of Perugia}
  \city{Perugia}
  \country{Italy}
}

\author{Giuseppe Liotta}
\email{giuseppe.liotta@unipg.it}
\affiliation{%
  \institution{Dept.~of Engineering, University of Perugia}
  \city{Perugia}
  \country{Italy}
}

\author{Falk Schreiber}
\orcid{0000-0002-9307-3254}
\email{falk.schreiber@uni-konstanz.de}

\affiliation{%
  \institution{Dept.~of Computer and Information Science, University of Konstanz}
  \city{Konstanz}
  \country{Germany}
}
\affiliation{
  \institution{Faculty of IT, Monash University}
  \city{Clayton} 
  \country{Australia}
}

\renewcommand{\shortauthors}{Kerle-Malcharek et al.}

\begin{abstract}
Immersive technologies, such as virtual and augmented reality, are transforming digital heritage by enabling users to explore and interact with culturally significant sites.
It is now possible to view and augment digital twins, or digitally reconstructed versions of them, and to enable access to previously unreachable locations for a broader audience.
Here, we investigate retrieval-augmented generation (RAG)-based avatars as an interface for accessing further information about digital cultural heritage objects while immersed in dedicated virtual environments.
We present a requirement design space that spans the application realm, avatar personality, and I/O modalities. We instantiate it with a RAG system coupled to a conversational avatar in a virtual reality (VR) environment, using the Maxentius mausoleum from the 4$^{th}$ century AD as a case study, through which users gain access to curated on-demand information of the digitised heritage object.
Our workflow utilises scholarly texts and enriches them with metadata. We evaluate various RAG configurations in terms of answer quality on a small expert‑crafted question-answer set, as well as the perceived workload of users of a VR setup using such a RAG avatar.
We demonstrate evidence that users perceive the overall workload for interacting with such an avatar as below average and that such avatars help to gain topical engagement.
Overall, our work demonstrates how to utilise RAG-driven VR avatars for archaeological purposes and provides evidence that they can offer a pathway for immersive, AI-enhanced digital heritage applications.
\end{abstract}



\maketitle

\section{Introduction}
Retrieval-augmented generation (RAG) systems have elevated large language models (LLMs) from mere text generation software to general-purpose systems capable of specific, diverse, and factual language~\cite{lewis2020retrieval}.
As a result, conversational agents can serve not only as a conversation partner, but also be utilised in several scenarios, such as a coding assistant or a digital expert for teaching, even in augmented or virtual reality (AR or VR)~\cite{10972698}. For example, they can be tuned for knowledge retrieval for cultural heritage (CH) purposes~\cite{zhang2025provenance}, or as conversational partners about archaeological topics in teaching scenarios, museums, and more.

Accessing CH data in AR or VR regarding the subject of three-dimensional (3D) reconstructions or contextual information poses a whole set of problems, in particular regarding the sheer quantity of the available data and their structuring, which is why conversational agents for archaeological topics are so promising. Those can synthesise information with only a limited amount of effort by the user.
Thus, an easy-to-use interface that could retrieve and summarise the data directly inside VR applications and even interact with it would have many benefits. It can serve as an interactive guide through the environment or synthesise knowledge. It could provide further information about different scholarly opinions or even present those in the form of different reconstruction hypotheses. 

Such an AI-driven VR conversational agent, while providing many benefits, also introduces new design challenges.
For example, how various implications of a system’s overall context can impact the agent's expected behaviour or in which ways the users' expectations towards the agent's level of expertise have to be accounted for.
Those considerations become even more relevant when considering that such agents, depending on the context, will be realised as an avatar in AR or VR.
Despite the complexity of such considerations, there is an increasing interest in such environments in digital heritage, which makes sense as they can help to foster data understanding through immersion and immersive analytics~\cite{marriott2018immersive,kyrlitsias2020virtual}.

Thus, with this work, we focus on enabling an AI-assisted exploration of visual CH data by positioning users inside a VR environment accompanied by a RAG-driven avatar.
Through positioning users in a digital version of a given monument, we allow them to visually experience the monument while being able to pose questions verbally to the RAG-driven avatar.
This integrates 3D reconstructions and historical context in one environment, while maintaining a low entry-threshold in terms of usability, which makes the overall concept flexible for multiple applications, whether it is tourism, research, or education.
To properly argue for such a system's flexibility, we propose a requirement space model for AI-driven avatars in VR for CH purposes that addresses various application realms.
As a practical application, we also developed a VR prototype  within our framework with an AI-driven avatar that will reply to questions with an authorial narrative style.

Our CH example for this work is the Maxentius mausoleum, a monument from the 4$^{th}$ century AD at the Via Appia Antica in Rome. 
The question of how this mausoleum might have looked during its prime remains unanswered.
There are two main hypotheses~\cite{Johnson2009,rasch1993mausoleum} based on monuments in relative proximity and of likely same typology, and in-situ evidence.
Through its rich history and the availability of 3D reconstruction models for this monument~\cite{kerle2025junction}, the mausoleum of Maxentius makes an excellent case study for our considerations.
In addition to prototyping our system, we present a full overview of the data and setup.

Finally, we evaluate the resulting RAG-LLM workflow by comparing different RAG configurations to determine how much augmentation is needed to meet the system’s objectives by measuring complementary metrics.
The evaluation metrics are composed of classic natural language processing metrics and of a LLM-as-a-judge metric to cater for not only the general semantic quality of the RAG's answers, but also whether the system can provide answers with a certain depth.
Furthermore, we conducted a NASA-TLX workload study with users who use our prototype to obtain answers to specific questions.
The user workload study aimed to determine if such avatars are sufficiently easy to use to be considered valuable additions in virtual environments. 
The major contributions of this paper are thus:
\begin{enumerate}
    \item A methodological approach to the design of an AI-driven avatar in the context of cultural heritage in immersive environments;
    \item a RAG-driven VR avatar prototype for the Maxentius mausoleum as a use case;
    \item an evaluation of various RAG configurations that quantifies the quality of its answers; and 
    \item a user study to evaluate the workload users have when using a system such as our prototype.
\end{enumerate}

The remainder of this paper is organised as follows:
Section~\ref{sec:related-work} provides an overview of related work.
Section~\ref{sec:design-methodology} introduces the three phases of the Avatar design process.
Section~\ref{sec:system-evaluation} discusses how we evaluated our system design to see the impact of our considerations in terms of answer quality and workload for users of a VR RAG avatar.
Section~\ref{sec:results} shows and discusses the results of both evaluations.
In Section~\ref{sec:discussion} we discuss our work, its limitations and future work.
In Section~\ref{sec:conclusions}, we conclude our work.

\section{Related Work}\label{sec:related-work}

\subsection{Immersive (VR/AR) Interfaces for Heritage Exploration}
When discussing the dissemination of archaeological findings, it is impossible to ignore the growing popularity of extended reality (XR) applications—an umbrella term encompassing VR, AR, and mixed reality (MR)—that enhance user experience.
This is a consequence of the strong benefit Virtual Archaeology enjoys from the strong sense of presence that immersive environments (IE) can evoke. Kyrlitsias et al.~\cite{kyrlitsias2020virtual} demonstrated this by building a 3D environment to explore the Choirokoitia site. Their feasibility study revealed a higher sense of presence in the virtual reality (VR) version of their reconstruction. Presence is a form of psychological immersion that also can be called engagement~\cite{Bueschel2018}, and is an important concept, as there is evidence that task performance and recall are boosted through heightened engagement~\cite{makowski2017being}.

For a museal environment, Ariya et al. went as far as to investigate which XR environments might be better suited for museums, concluding that VR and mixed reality (MR) provide great potential to increase user engagement~\cite{DBLP:journals/vr/PakineeWWIP25}.
AR has also seen over a decade of varying uses in 3D exploration of cultural artefacts, virtual museums~\cite{Spallone2024}, education, tourism and more, as Boboc et al.\,present in their survey~\cite{boboc2022augmented}.

By the nature of the topic of cultural heritage (CH), citizen science is as much a part of it as are applications.
For instance, virtual museums can benefit from having multiple people participate, as shown by Ibanez et al., who created a system to further the comprehension of what a Roman domus is through a multi-user experience~\cite{hernandezEzAl}. The general public can participate in data acquisition for the creation of 3D reconstructions and experience the required work before a reconstruction is produced~\cite{6744726}.

Another concept that can enhance immersion in XR applications is avatars~\cite{10049669}, as they can introduce a way to interactively communicate with the system itself, or even with non-playable characters driven by LLMs. Those special avatars can be chosen to be diegetic in-universe characters, distinctly increasing the presence of users, according to Nielsen and colleagues~\cite{NielsenNMON25}.

In XR, VR specifically stands out as a technology with a set of multimodal and multisensory capabilities.
Multi-modality refers to providing users with ways to interact with the environment beyond a simple point-and-click paradigm, such as gestures and speech. Multi-sensory refers to the utilisation of various user senses apart from the usual visual and auditory channels, to convey information, as research shows that combining unconventional senses with standard audiovisual feedback can be beneficial~\cite{cooper2018effects}.

In addition to outreach, immersive media are increasingly employed for scientific knowledge acquisition, such as investigating how to properly produce 3D models of a monument for VR environments~\cite{rodriguez2024systematic}, the investigation of lighting conditions~\cite{happa2009virtual}, discussing methodologies for virtual restoration~\cite{Ma25}, or comparing different reconstruction hypotheses~\cite{kerle2025junction}. Meanwhile, methodological research is also part of the field, with researchers like Doerr identifying different sub-genres that developed in virtual archaeology~\cite{doerr2009ontologies}.

Such a distinction is beneficial in terms of use-case considerations. For example, a system focusing on research can gain a lot from the field of immersive analytics, where researchers use embodied tools to increase data understanding~\cite{marriott2018immersive}. Immersive tools are capable of increasing information recall~\cite{mark2003there} and supporting reasoning in general~\cite{elsayed2015using}, while Ragan et al.\,also report gains in spatial‑judgment accuracy when participants explore a monument virtually rather than through static images~\cite{ragan2012studying}.

Considering all the discussed aspects, it is clear that much of the research work involves XR technology for CH-related topics, with the clearest distinction and highest potential being research and dissemination.

\subsection{Conversational Avatars, Retrieval‑Augmented Generation for Cultural Heritage}
Parallel to the development of XR systems related to the CH context, large language models (LLMs) see increased use, too. Trichopoulos, for instance, discussed in his work how a LLM could be used as a virtual guide in museums in the future~\cite{trichopoulos2023large} and discusses speech-to-text (STT) and text-to-speech (TTS) setups for this purpose. In their recently published work, they further evaluate the potential of LLM chatbots and point out that future developments should involve RAG-driven systems~\cite{trichopoulos2026evaluation}. Bongini et al. also propose a digital guide for museums, but to ask questions and retrieve visual content with relevant information, rather than having a conversation~\cite{Bongini2020}. Many of these works, however, do not emphasise domain-knowledge-driven systems whose focus lies in participating experts not only in training the data, but also in designing these systems.

Due to their ability to retrieve their knowledge from a curated knowledge base before generating an answer to a query, RAGs have been widely used already as conversational experts in recent developments~\cite{10972677,zhang2025provenance}, which is especially interesting for considerations about virtual guides. RAG models can be used to create LLMs whose knowledge stems from a steered and curated knowledge base~\cite {lewis2020retrieval}, directly or indirectly quality-controlled by developers and curators alike. The former essentially ensures that the knowledge fed to the RAG is appropriate for the use case. The latter can utilise one of the metrics that exist to evaluate RAGs as conversational agents~\cite{miyaji2025evaluating,zhang2019bertscore}.

However, it is much harder to say if an avatar actually fulfils any sort of quality criteria and what those criteria are. One way to ensure a certain level of quality has been developed by researchers such as Rashik et al., who propose design considerations for digital avatars~\cite{rashik2024beyond} which they summarise in a table. That table consists of elements that thematise visual appearance, target system, or type of device, among others. With their work, they provide a strong topical survey that categorises 266 publications based on their content into subjects related to designing an avatar.

Their focus lies on what has already been created in terms of software solutions, however, and less on the types of different considerations that could flow into a conversational avatar from an interdisciplinary perspective, such as archaeology. In this regard, Barnett et al. provide a list that shows typical pitfalls to avoid when designing a RAG ~\cite{barnett2024seven}. Optimisation of RAGs can help to improve the outcome, and there already exist works that address optimisation techniques~\cite{jiang2025rago}. For instance, metadata enrichment can be utilised for re-ranking and other answer steering processes~\cite{poliakov2024multi}.
This can be utilised as filters, for specific authors, for instance, or for different statements about an object to compare different hypotheses.
This tagging can be realised through the aforementioned metadata enrichment, or through graph‑based retrieval~\cite{peng2025graph}.

\section{Design Methodology}\label{sec:design-methodology}
\begin{figure}
    \centering
    \includegraphics[width=.6\linewidth]{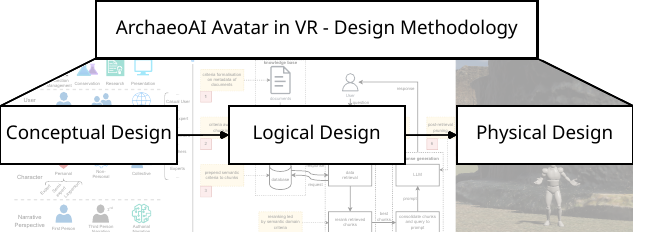}
    \caption{This illustration shows the three overarching steps taken to design and implement a digital AI-driven avatar for immersive spaces in an archaeological context. The conceptual design is a means of deciding what constitutes the requirements for the system, the logical design to decide on the overall architecture, and the physical design to do the actual implementation.
    }
    \label{fig:3step}
\end{figure}

The primary goal is to provide an immersive space utilising VR for exploring archaeological objects with the assistance of a digital expert avatar. In our specific case, this object is the mausoleum of Maxentius at the Via Appia Antica in Rome.
There are plenty of ways to achieve this, and we want to provide the means to approach this in a structured way.
We therefore propose an overall design that encompasses three overarching topics~(see Figure~\ref{fig:3step}): (1) the conceptual design with the goal to gather the systems overall goal, requirements and the related user tasks; (2) the logical design, where the output of the conceptual design is used to design the system architecture; and (3) the physical design, which corresponds to the systems actual implementation, i.e.\,actual code, models and so on.

\subsection{Conceptual Design}
\begin{figure}
    \centering
    \includegraphics[width=.7\linewidth]{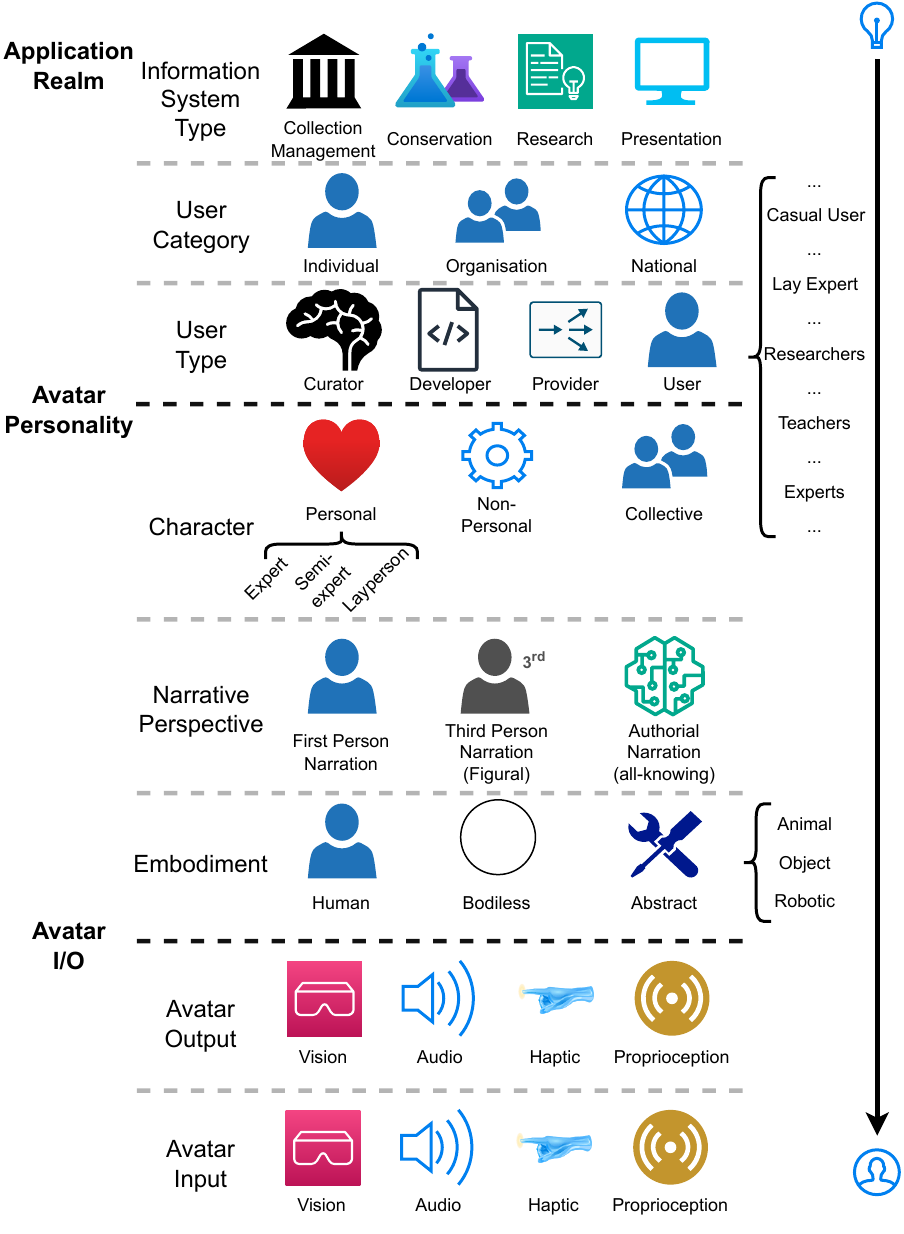}
    \caption{This illustration shows our proposal for a requirement space for avatars in virtual archaeology applications. It contains three main building blocks: (1) The application realm indicates who the final users of such a system are. (2) The avatar personality helps to figure out what type of answers a user could expect from a given avatar. (3) The avatar I/O thematises the need to decide on how input is accepted by an avatar and how it provides information, given the focus on immersive environments. The design spaces (system type, user category, $\dots$) refer to the following publications, in order from top to bottom: Doerr~\cite{doerr2009ontologies}, Deshpande~\cite{deshpande2022responsible}, our suggestion with the "User" based on Walsh~\cite{walsh2016user}, Bartsch~\cite{bartsch2025epistemic}, van Peer~\cite{van2001new}, Rashik~\cite{rashik2024beyond}, and Martin~\cite{martin2022multimodality} for I/O.}
    \label{fig:requirementSpace}
\end{figure}

For the conceptual design, we propose three building blocks that are interchangeable, depending on the development circumstances (see Figure~\ref{fig:requirementSpace}).
Those building blocks are the application realm, the avatar’s personality, and the I/O between the avatar and the user.
With this, we determine for whom the avatar is designed, how it is supposed to be perceived and act, and how this can be realised through the means of VR-related in- and output considerations.
In the centre of the image are eight rows, each of which constitutes a major design space that has to be considered by system designers.
Each row contains three or four directional concepts, symbolised by icons and a descriptor that a system designer can choose.
These concepts represent possible choices in the topical design space, where the choice of one narrows the overall design space for the system itself.
Some concepts have further sub-categories, which are denoted in the figure.
The three building blocks are separated by thick black dotted lines. The major design spaces are separated by thinner grey dotted lines.
All concepts were chosen from peer-reviewed literature, with a few exceptions, described later in the chapter. 

The black arrow on the right side illustrates that we consider the image to show a process, from idea (lightbulb) to avatar (human head).
System designers choose, row by row, one (or multiple) option(s) per concept to determine the requirements for the final system that is being designed.
Therefore, the overall illustration can be considered to show how to determine one's own system requirement space.

At the end of the conceptual design, system designers decided on a specific set of tasks that users have to fulfil in order to interact with the system.
For example, they have to verbally pose questions that can range from very general to very specific.
Then, the avatar will forward the interaction parameters for processing to the RAG.
This RAG processes the interaction (e.g. asked questions), sifts through its knowledge base to create a fitting answer and responds with the answer, which is then relayed to the user through the avatar (e.g.\,text-to-speech (TTS)).
It refuses to answer if it can't find fitting information in its database.

\subsubsection{Application Realm}
To formally describe the application realm, it is necessary to understand what the system is built for and in which context.
For this, we refer to Doerr's work, who identified four key sub-topics in virtual archaeology as an example ontology~\cite{doerr2009ontologies}: (1) Collection management, such as exhibitions; (2) conservation work, like using solutions for actually conserving CH objects; (3) research as a knowledge gain topic; and (4) presentation in the form of teaching.
With this, we can already narrow the scope for the  kind of consumers of the final system. 
On a higher level, it is also required to consider the category of users. According to Deshpande et al., those might be (1) individuals, (2) organisations, or even (3) nations~\cite{deshpande2022responsible}.
For most cases, individuals will likely be the answer.
On a further level, we suggest that a classification of who is operating the system is made, while differentiating between (1) curators, (2) developers, (3) providers, and (4) consumers, which we simply refer to as users.
Curators are the domain experts who can steer the narrative direction of the avatars.
Developers maintain the system on a technical level. 
Providers are those who are responsible for the system as such, and who take responsibility for issuing updates, making it available and taking action if the system shows faulty behaviour. 
Users of the system are characterised in Walsh's survey~\cite{walsh2016user}.

So, after making those selections, system designers decide on a very specific set of users and, thus, very specific tasks that the system has to service and system designers have to consider.

\subsubsection{Avatar Personality}
With the application realm specified, the personality of the avatar needs to be defined.
This is important for understanding the kind of content the RAG requires for its knowledge base and the structure that data requires.
In a museum, a user could expect a guide-like figure or some information source without a resemblance to a human-like personality.
Bartsch et al.~distinguish three main categories of such knowledge persona types: (1) personal, (2) non-personal, and (3) a collective of characters, like academies or boards~\cite{bartsch2025epistemic}.
They refer to them as epistemic authority, and by deciding on one character type, system designers choose the depth and type of knowledge the avatar has to have available.
Then there is the question of the narrative perspective, as the avatar could be describing something from (1) a first-person or (2) a third-person view, or maybe (3) an all-knowing, authorial type altogether~\cite {van2001new}.
Lastly, the visual representation of the avatar needs to be considered.
Rashik et al.~name (1) human-like, (2) animal-like and (3) robotic as main embodiments~\cite{rashik2024beyond}.
We reshape their classification by saying, we either have (1) a human-like avatar, (2) a bodiless one, or (3) an abstract one, which encompasses everything that is neither (1) nor (2), such as animals, objects, or robots.

The goal of designing an avatar that adheres to such character traits is to actively address the human perception of virtual CH environments and to fulfil their expectations towards such environments. It provides the possibility to exploit a number of immersion-enhancing factors, thus positively affecting the value digitised CH objects can have for the targeted user groups.

\subsubsection{Avatar I/O}
The third building block for our requirement space model is to decide on the avatar's input and output of information.
There are numerous potential resources on what types of input and output one could consider for VR applications.
We only focus on the most prevalent ones: (1) visuals, (2) audio, (3) haptics, and (4) proprioception~\cite{martin2022multimodality}.
The user can see the avatar, if not designed to be bodiless, and hear it, if it is designed to provide audible output.
It is possible to relay information via haptic feedback, and the combination of different senses can even be utilised to induce some fake sense of spatial awareness, which, in turn, helps to further immerse users.
On the other hand, it is possible to use multimodal means to interact with an avatar. This involves speaking to the avatar, pointing at objects, or using one's own virtual proximity to a virtual object.

By selecting I/O capabilities for the avatar, system designers influence how appropriate the interaction with the avatar is perceived by users for the task they have, and how well they perform. For instance, if the avatar is supposed to be a conversation partner, the user would expect to be able to talk to it in the first place. Further conversational factors, such as gestures, also account for I/O considerations with more or less relevance depending on the targeted audience (e.g.\,mute people).

\subsubsection{Our Design Selection}
\begin{figure}
    \centering
    \includegraphics[width=0.9\linewidth]{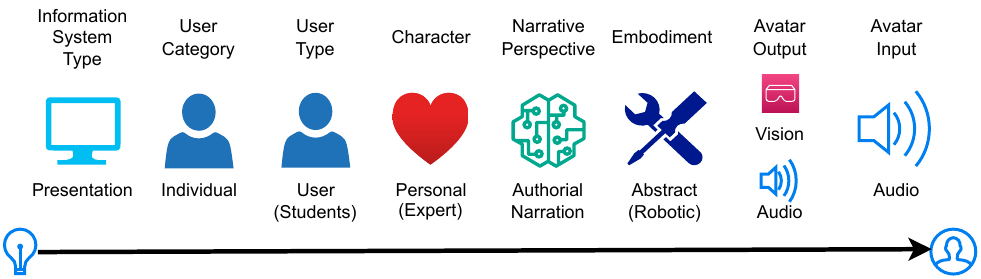}
    \caption{An illustration of the selection for our setup for the Maxentius mausoleum use case, for the elements refer to Figure~\protect\ref{fig:requirementSpace}. Our application realm is chosen to serve (semi-)expert individuals for research purposes. The avatar personality of the avatar is chosen to be personal-expert with an authorial narration style and an abstract-robotic embodiment. The output for the avatar is vision through simplified gestures and audio. The input is audio. The left-to-right-arrow is indicating that the choices made are now horizontally displayed, instead of the vertical selection order as proposed in the original picture.}
    \label{fig:selection}
\end{figure}

For our current prototype, we chose our application realm to be presentation for individual users with a learning incentive, so students, or lay experienced users.
The personality of the avatar is supposed to be personal on a semi-expert to expert level with an authorial narration style, and have an abstract robotic embodiment.
The I/O for the avatar is mainly audio-audio at this point, even though the avatar has minimalistic body movements to indicate processing of the question to the RAG (see Figure~\ref{fig:selection}).

Since the avatar is supposed to serve student-level users in a presentation context, it is arguably required to have well-curated works that directly relate to the object of interest and serve as a knowledge base.
Due to the low amount of data available for most archaeological objects, this poses a challenge. 
Especially because, due to the avatar’s (semi-)expert level and authorial narration style, the avatar should be restricted to deny answers that are not founded, and it should not be allowed to have high levels of creativity, meaning a low temperature for the QA retrieval chain.

\subsection{Logical Design}
The logical architecture of our system consists of the VR application, a local database, and the RAG running locally.
The VR application records the user's utterances and uses speech-to-text (STT) to transcribe them. If those utterances are recognised as questions, they are transmitted to the RAG, which uses the transmitted question as a query.
The RAG’s retriever utilises the embedding that is stored on the local vector database.   
The RAG’s response is returned to the VR application, where it is translated to spoken language using TTS.

\begin{figure}
    \centering
    \includegraphics[width=.8\linewidth]{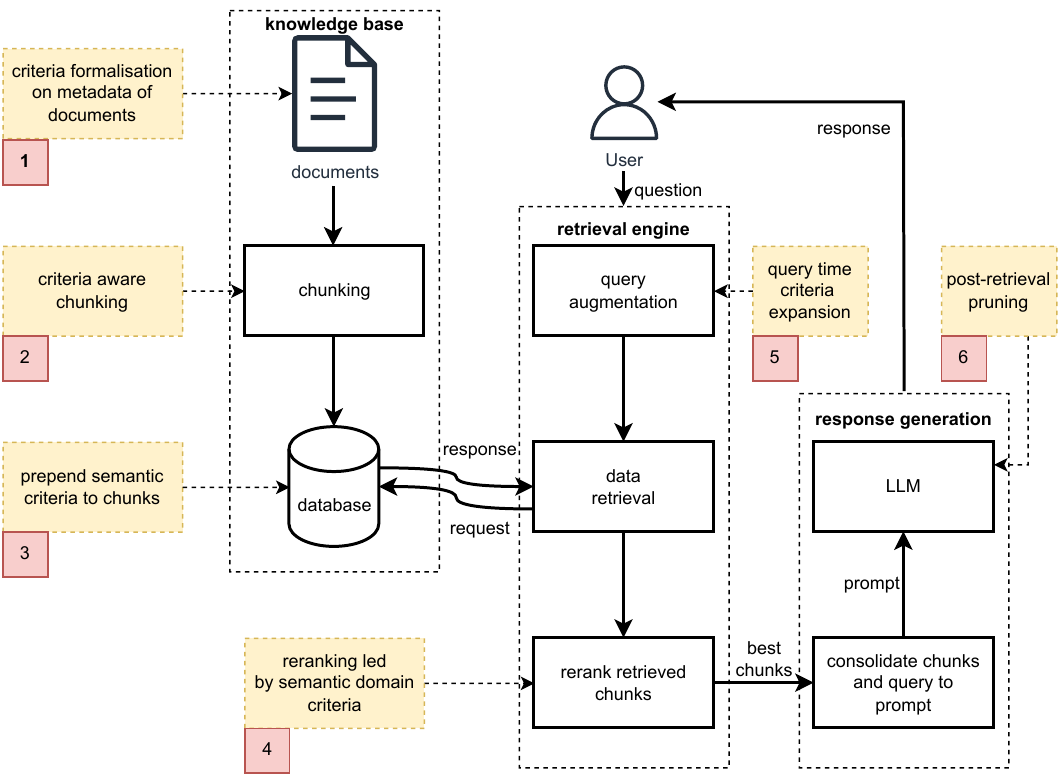}
    \caption{Simplified general workflow of a RAG. In yellow are a couple of identified ways to include our criteria.}
    \label{fig:rag}
\end{figure}

Inside the RAG, we can utilise several strategies to use domain knowledge to steer the answers of the avatar. Curating such a system without extended technical expertise, e.g.\,curation by archaeologists without AI expertise, limits the reasonable possibilities of meaningfully tuning a RAG. In Figure~\ref{fig:rag}, we show ways we consider reasonable to incorporate domain-specific reasoning. Those ways to incorporate domain knowledge and domain-specific reasoning are inspired by Barnett et al., who identified seven failure points in RAG engineering~\cite{barnett2024seven}, and are marked with pink numbered squares, and mainly utilise either metadata or prompt engineering.
It is possible to augment the data fed into the knowledge base~(1) and how they are structured~(2), how the ranking of the chunks is determined upon querying~(3), the chunks that are part of the final response~(4,5), and the query that is ultimately fed into the LLM~(6).
To emphasise different hypotheses from varying sources, for example, a simple way is to utilise query time criteria expansion~(5), by adding to each prompt to look for contradicting statements and pointing them out.
Another way to try to achieve hypothesis emphasis is to employ criteria formalisation on metadata of documents~(1), by building an ontology-based knowledge graph. For example, by using CIDOC-CRM~\cite{oldman2014cidoc} as an ontology, for each document, query for potential hypotheses and add a marker to the metadata to do reranking led by semantic domain criteria~(4) afterwards.

After the interaction of the user is recognised by the VR system, it is relayed to the retrieval engine of the RAG.
There, it is subject to retrieval augmentations and utilises the knowledge base to obtain an answer.
After the system has retrieved the highest ranking chunks, it triggers a LLM response that is relayed to the VR system, where it, in turn, gets processed to create a proper response to present to the user.

\subsection{Physical Design}
The VR environment is created in Unity with C\#. We used the Maxentius mausoleum and a simple avatar representation as our environment, see Figure~\ref{fig:teaser}.
When the user asks a question while situated inside the VR environment, an STT, implemented using Evgrashin 's~\cite {macoron_whisper_unity} implementation of  the OpenAI whisper STT framework~\cite{radford2023whisper}, will translate the question into a query and relay it to the locally hosted RAG system.
The local RAG system is realised using FlowiseAI~\cite{flowiseai_flowise} and consists of five nodes, an Ollama embedder, an Ollama chat model, a Qdrant database as a knowledge base, a buffer window memory with k=2 for only being able to rephrase questions, and a conversational retriever QA chain as a conversation manager.
Once the final answer is retrieved, it gets sent back to the VR environment.
There, TTS using the Piper framework~\cite{hansen2023piper}, transforms the query answer of the RAG into audible speech that the user can perceive.

To adhere to the necessity of well-curated literature as a basis for our knowledge base, we used only peer-reviewed publications that directly or indirectly address the mausoleum of Maxentius in Rome.
To enrich the metadata used during retrieval, we manually added general information, such as author names, titles, and the type of literature. 
Furthermore, we added a subjective criterion, namely the relevance of the literature in question.
First of all, we consider the work of Jürgen Rasch about the mausoleum~\cite{rasch1993mausoleum} to be the main work to consider.
Then, we classify all further works that were used and focus on the mausoleum as relevant.
Lastly, we classify as adjacent those works that mention the mausoleum, but focus on related topics, such as the emperor Maxentius or Roman history. 
With such a field, curators can directly influence and steer the system's output, as discussed before, via semantic information, such as their own assessment of what constitutes the most pertinent piece of literature about a given CH object.
Further influence on the metadata, such as adding possible different hypotheses, would add a much higher complexity at this point of contemplation, which is why we refrain from incorporating this in the current prototype.
Thus, in the end, our final metadata has the following structure:

\begin{itemize}
    \item Author
    \item Title
    \item Publication type
    \item Relevance: \emph{main}\,|\,\emph{relevant}\,|\,\emph{adjacent}
\end{itemize}

To prepare the text for further usage, we applied a basic preprocessing toolchain, first processing the PDF files to plain-text via OCR, using a Python wrapper for Tesseract OCR~\cite{smith2007overview}.
The plaintext then has undergone additional basic cleaning (such as removal of empty lines, etc.) before ingestion into the knowledge base.
The embedder for our prototype is MLE5~\cite{wang2024multilingual} and the chat model Llama 3.1~\cite{grattafiori2024llama}.
The temperature of the system was set to 0.3.
The database we are using for our embedding is Qdrant, as it offers a higher versatility, also in terms of metadata payloads~\cite{ozturk2024performance}, and the corresponding vector retriever. 
The top k we chose to be 4, after testing different setups and finding that k=4 is the best setting for our purposes in terms of answer quality and retrieval time.

\begin{table}[h]
\centering
\begin{tabular}{|l|l|l|c|c|}
\hline
Name      & ShortName & Category & Parameters&citation\\ \hline
Multilingual-E5-large-instruct&MLE5& Embedding&0.6B&\cite{wang2024multilingual}\\ \hline
Qwen3-embedding&Qwen3& Embedding&4b&\cite{zhang2025qwen3embeddingadvancingtext}\\ \hline
Llama 3.1&Llama3.1&Chat&8b& \cite{grattafiori2024llama} \\ \hline
teuken-7b-instruct-research&teuken&Chat&7b&\cite{ali2025teuken7bbaseteuken7binstructeuropean}\\ \hline
Llama-3.3-70b-instruct&Llama3.3&Chat&70b&\cite{grattafiori2024llama}\\ \hline
\end{tabular}
\caption{This table shows the different models we used for our metrical evaluation assessments. The first column is the name of the model, and the second column is a shorthand for their names. The third column shows in which context the models were used. The fourth column addresses the number of parameters of those models, and the last column contains the reference to the respective models.}
\label{tab:models}
\end{table}

The data were additionally processed in order to create and test a knowledge graph as an underlying knowledge base. This step was done using the LLM-Graph-Builder application \cite{llm-graph-builder} of Neo4j Labs \cite{neo4jlabs}. The process involved the same LLM embedder and chat models as above for the chunks embedding and automatic extraction of the entities and relationships of the knowledge graph. This phase was augmented with a prompt asking to employ a CIDOC-CRM-driven extraction.
Since we were only prototyping, we decided not to apply further manual enhancement and post-processing phases like graph schema definition or duplicate nodes merging on the generated knowledge graph.
The RAG (in this case commonly addressed as a GraphRAG \cite{wang2025graph}) uses the Vector+Graph+FullText retrieval approach available in LLM-Graph-Builder that combines the text retrieved from similarity search on vector index, keywords search on full text index and Cypher queries to the graph database.

\section{System Evaluation}\label{sec:system-evaluation}
The restrictions of our setup, namely, concise formulation, refusal to answer incorrectly or off-topic, low creativity, and the utilisation of a strictly controlled set of publications, necessitate an evaluation of the quality of the answers. Furthermore, as the system is an avatar that interacts with a user in VR, we want to measure the actual workload of users who use such a system. To do our evaluations, we generated 10 question-answer (QA) pairs, with questions related to the mausoleum of Maxentius and answers whose quality is semi-expert to expert level in the subject matter, with the assistance of domain experts of the Maxentius Mausoleum and adjacent topics. Those QA pairs serve as a ground truth for our experiments.

\subsection{Metrical Answer Quality Assessment}
To evaluate how different domain-knowledge-driven steering strategies impact the quality of our output, we decided to compare different setups of the same RAG system.
In Table~\ref{tab:models} we list all the models used for the experiments and their size.
The main criteria for the choice were model size and multilingual ability. Smaller models allow easier, cost-effective on-premise usage and prove more adequate for real-time or near real-time applications like ours; additionally, for comparison purposes, we included a setup with bigger embedding and chat models, namely Qwen3-embedding and Llama3.3.
The temperature for all the LLM models was set to 0.3, as a lower temperature leads to less interpretative chatter from the LLM.

For all setups, we use the same metadata attached to the chunks, as described before, with author, title, publication type, and relevance.
Those are utilised during query time, as criteria expansion to bias the system without inhibiting it from reaching out to less related chunks.

The first setup is a simple RAG as seen in Figure~\ref{fig:rag}, where the data has been ingested into the knowledge base with a basic recursive character splitter with a chunk size of 1000 and an overlap of 200.
This setup serves as a ground truth for the evaluation scheme we discuss in this section.
The second setup has the same chunks, where the metadata has been augmented with the relevance of the corresponding document.
The third system, instead of relevance, enhances the chunks with the knowledge graph.
The first two setups have also been varied with the usage of different LLM models for the embedding and generative phases, leading to a total of seven different setups.

For the evaluation, we adapt the evaluation scheme as seen in the work of Miyaji et al., who use complementary evaluation metrics to achieve a more complete picture of the performance of conversational agents~\cite{miyaji2025evaluating}.
We use a two-fold evaluation methodology, where we use LLM as a judge as well as traditional NLP metrics. 
For this, we utilise the 10 QA pairs that form our ground truth.
The ground truth questions were then answered by the RAG, and we evaluated its answers with the help of the evaluation scores as seen in Table~\ref{tab:scores}.

\begin{table}[]
\begin{tabular}{|l|l|}
\hline
Score       & \multicolumn{1}{l|}{Description}          \\ \hline
1           & \begin{tabular}[c]{@{}l@{}}The response is incomplete, factually incorrect, or irrelevant to the user’s\\ query, potentially leading to misunderstanding or misinformation.\end{tabular} \\ \cline{1-1}
2           & \begin{tabular}[c]{@{}l@{}}The model attempts to answer but provides partially incorrect or vague\\ information, with significant omissions or lack of clarity.\end{tabular} \\ \cline{1-1}
3           & \begin{tabular}[c]{@{}l@{}}The model offers a generally accurate and relevant response, but may lack\\ full detail or leave some aspects of the user’s query unaddressed.\end{tabular} \\ \cline{1-1}
4           & \begin{tabular}[c]{@{}l@{}}The response is factually sound and covers most key points with clarity,\\ though there may be minor gaps in completeness or nuance.\end{tabular} \\ \cline{1-1}
5           & \begin{tabular}[c]{@{}l@{}}The model delivers a comprehensive, precise, and clearly articulated\\ answer that fully addresses the user’s query with high factual integrity\\ and helpful context.\end{tabular} \\ \hline
\end{tabular}
\caption{The scoring table we used for the LLM-as-a-judge evaluation, as seen in the work of Miyaji et al.~\cite{miyaji2025evaluating}}
\label{tab:scores}
\end{table}

The traditional NLP metrics we used are METEOR~\cite{banerjee2005meteor} as a pure lexicographical metric and BERTScore~\cite{zhang2019bertscore} as a metric for semantic similarity to the respective expert answers of the question-answer-pairs.
As pointed out in the literature, METEOR suffers from shortcomings common to most NLP-based metrics; nevertheless, for comparison reasons, we provide such results.
Regarding BERT-based metrics, one key factor is the quality of the embeddings employed to measure the semantic closeness between chunks; fine-tuned models prove more adequate to capture the specificities of a domain, providing more refined embeddings for specialised terminology.

The LLM as a judge system we used was Prometheus2~\cite{kim2024prometheus}, which Miyaji et al. refer to as a state-of-the-art fine-tuned LLM evaluation model, which would rate the two answers from the respective RAG systems according to the same 1-5 rating scale used in the original publication.
The criteria taken into account by the LLM must be carefully selected for a fair scoring process, and are application dependent; while for pure conversational or support tasks factors such as fluency and helpfulness could prove more adequate, for our purposes it is paramount that the system focuses on retrieving exact information, presents it coherently with the sources and does not in any case distort scientific facts and hypotheses.
Therefore, in the judging process, we instruct the LLM to emphasise relevance, coherence and factuality, with the following prompt: "You are a helpful and precise assistant for checking the quality of the answer. Please rate the helpfulness, relevance, accuracy, and level of detail of the response."
According to \cite{wei2025systematicevaluationllmasajudgellm}, choosing a low temperature for the Judge LLM is important to guarantee consistency in the evaluation process. Therefore, we choose a value of 0.1; still, to account for the residual variability of the judgment, we run 15 evaluations and report the average numerical score.
\subsection{User Workload Study}
To evaluate the perceived workload associated with the interaction with our VR-LLM system, we conducted a user study employing the NASA Task Load Index (NASA-TLX), a widely adopted and well-established subjective workload assessment instrument~\cite{hart2006nasa}. The questionnaire used a discrete 1–21 scale for each dimension, where higher values indicate higher perceived workload.
For the study, we used the 10 question-answer pairs we obtained from domain experts and gave the users the task to find the answers to the questions with the assistance of the avatar.

\begin{figure}
    \centering
    \includegraphics[width=.6\linewidth]{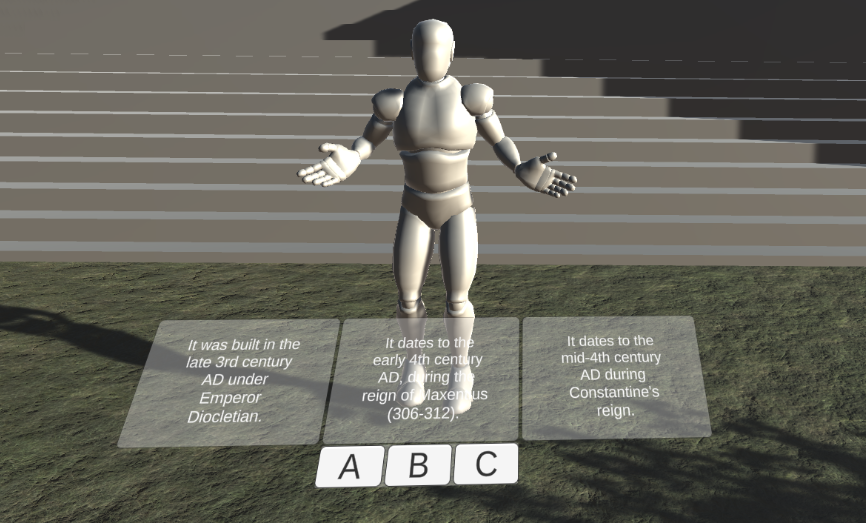}
    \caption{\label{fig:studyDemo}This image shows an impression of what users saw during the user study. The grey fields show the potential answers users could choose from, in this case, for the question "What is the dating of the mausoleum?"}
\end{figure}

During the study, users stood in front of the avatar, positioned within a reconstructed virtual environment of the Maxentius mausoleum. A small interface window displayed the question they were asked to find the answer to. After verbally posing the question to the avatar and listening to its response, participants were presented with three possible answer options. One corresponded to the shortened expert-validated answer, while the other two were generated to appear plausible but were intentionally incorrect. Shortened answer variants were used to avoid potential bias related to response length, as participants were required to carefully read and compare the alternatives before making a selection. Responses could be provided either by clicking the corresponding option or by verbally uttering “A”, “B”, or “C”. Both the order of the 10 questions and the placement of answer options were randomised for each participant (see Figure~\ref{fig:studyDemo}).

The procedure began with an instruction phase, during which participants were introduced to the study setup, the evaluation sheet, and the system interaction mechanics. This was followed by a short familiarisation phase within the virtual environment to allow participants to become comfortable with the avatar interaction. Subsequently, participants completed all 10 question–answer trials. At the end of the session, they filled out the NASA-TLX questionnaire to provide their own assessment of the perceived workload they had over the session. Participation concluded immediately after the questionnaire. To recreate the study, please refer to our Zenodo repository~\cite{kerle-malcharek2026}, where we provide all resources needed to do so.

\section{Results}\label{sec:results}
\begin{figure}
    \centering
    \includegraphics[width=.6\linewidth]{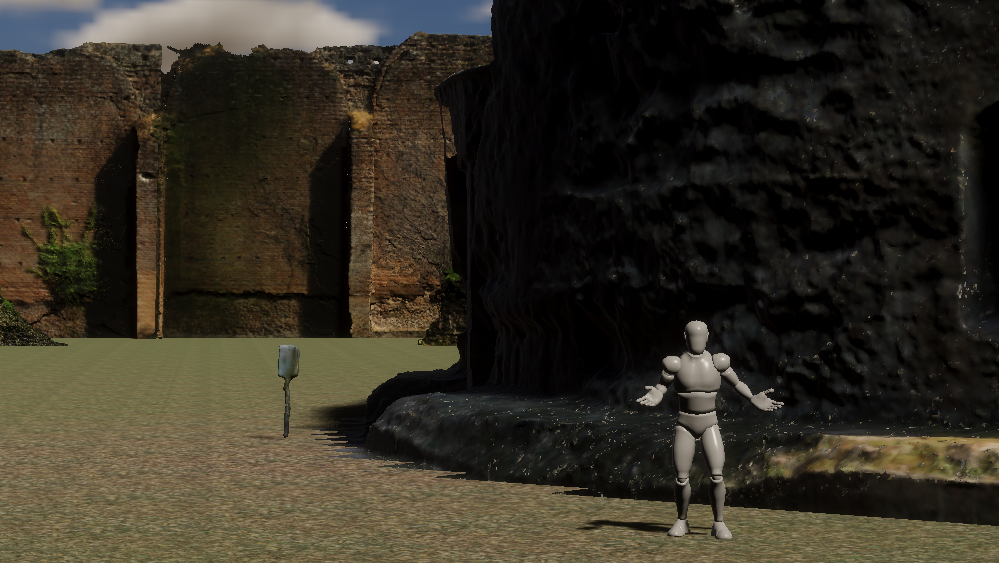}
    \caption{This is a teaser image of the final prototype environment. We can see the avatar standing in front of the digitised monument, while it is gesturing as an indication of it talking, thus responding to a question.}
    \label{fig:teaser}
\end{figure}

We created a VR environment with an avatar that users can talk to and that will answer questions related to the historical object inside that environment. The environment we chose is a digital version of the mausoleum of Maxentius as both, digitised version and reconstructed version, as seen in Figures~\ref{fig:studyDemo} and \ref{fig:teaser}. 

\subsection{Metrical Answer Quality Assessment Results}

\begin{table}
\centering
\begin{tabular}{|l|l|l|c|c|c|}
\hline
Embedding&Chat&Metadata/Metric&Meteor&BERTScore (F1)&LLM judge\\ \hline
MLE5& teuken& No relevance&0.272&0.779&2.67\\ \hline
MLE5& teuken& Relevance&0.260&0.768&2.73\\ \hline
MLE5& Llama3.1& No relevance&0.231&0.773&2.72\\ \hline
MLE5& Llama3.1& Relevance&0.223&0.778&3.13\\ \hline
MLE5& Llama3.1& Knowledge Graph&0.267&0.783&3.03\\ \hline
Qwen3& Llama3.3& No relevance&0.251&0.779&3.29\\ \hline
Qwen3& Llama3.3& Relevance&0.232&0.781&3.42\\ \hline
\end{tabular}
\caption{Metric evaluation results. The scores are averaged over 10 question/answer pairs. The first three columns show the setup that was tested, with the respective embedder, chat, and metadata used. The fourth and fifth columns display the METEOR and BERTScore scores per setup, respectively, and the last column shows the LLM-as-a-judge scores averaged over 15 runs. The complexities and full model names can be seen in Table~\ref{tab:models}.}
\label{tab:resultssetups}
\end{table}

The results in Table~\ref{tab:resultssetups} show our quantitative assessment of the answer quality. 
First of all, the traditional metrics METEOR and BERTScore show rather small variations, especially compared to the LLM-as-a-judge, so we focus more on the latter for evaluation.
The quality of the answers of the RAG, according to the score provided by the LLM-as-a-judge, consistently increases with model complexity, which was to be expected. 
More importantly, the score also improves with the added metadata and increasingly so if the underlying model gets more complex.
Across all setups, Qwen3 embeddings with Llama3.3 LLMs achieved the highest performance, in particular together with included relevance metadata, resulting in a 3.42 average score. 
Notably, the GraphRAG setup also caused an improvement, although for a much higher effort to set up compared to the applied metadata strategy, which even outperformed the GraphRAG. However, the GraphRAG employed here is only a comparably generic setup, and more sophisticated implementations will likely be more competitive. In particular, using a more context-specific prompt or a tailored graph schema for the entity recognition may lead to interesting improvements, but at the cost of an increasing management effort for non-expert users. 
Summarising, we can deduce that the system provides users with correct answers that might lack in depth with respect to the answers given by the expert. Therefore, the prototype could already be suitable for presentation and teaching uses, but would require some more refinements to address the needs of expert users.
As a last remark, the low variation in METEOR and BERTScore compared to the LLM-as-a-judge evaluation reinforces the importance of the latter for domain-sensitive assessment.

\subsection{User Workload Study Results}
\subsubsection{Workload Assessment}
Perceived workload was computed using the Raw NASA-TLX scoring procedure, defined as the unweighted mean of the six subscales (Mental Demand, Physical Demand, Temporal Demand, Performance, Effort, and Frustration). The questionnaire was answered once at the end of the session to obtain a global evaluation of the interaction with the avatar.
We conducted the user study with $n=15$ participants. The mean values per subscale, as well as the standard deviations, and the minimum and maximum values are shown in Table~\ref{tab:nasatlx}.  

\begin{table}
\centering
\begin{tabular}{lcccc}
\hline
Dimension & Mean & SD & Min & Max \\
\hline
Mental Demand & 11.87 & 4.60 & 4 & 20 \\
Physical Demand & 2.67 & 1.91 & 1 & 8 \\
Temporal Demand & 8.53 & 5.08 & 1 & 16 \\
Performance& 9.00 & 4.36 & 2 & 19 \\
Effort & 8.67 & 4.84 & 2 & 19 \\
Frustration & 9.47 & 6.78 & 1 & 21 \\
\hline
TLX Total & 8.37 & 2.75 & 3.83 & 11.50 \\
\hline
\end{tabular}
\caption{\label{tab:nasatlx}The mean and standard deviation results of the NASA-TLX with a 1-21 scale and $n=15$ of our user study.}
\end{table}
With the midpoint (mp) of our scale at 11, the mental demand, with 11.87, is slightly above mp, which can be attributed to the fact that the participants had to adjust to a very specific topic and had to deal with domain-specific vocabulary. Furthermore, if the avatar was not immediately responding in a way that the user might have expected, they had to adjust the way they posed the question.
Since the score is only slightly above mp, we can interpret this as meaningful cognitive engagement. 
The score for the physical demand is very low, with 2.67, which was to be expected given that the task only required users to talk to a digital avatar and wear a headset. The standard deviation here is also the lowest, indicating that the physical demand is generally understood to be low by all participants.
The temporal demand is smaller than mp, indicating that the response and interaction pace is perceived as acceptable, which is relevant if such a system is supposed to support researchers during reasoning and discussions.
Performance is below mp, hinting that users felt like the system could actually help them accomplish the task of answering questions.
The effort is also below mp, hinting towards cognitive engagement instead of distress, which is consistent with the mental demand observation.
The frustration score is also below mp and below mental demand, which indicates that the interaction is working for a majority of cases. Of course, it can not be ignored that the score is still in a range where it may either correlate with higher but unexpected engagement or faulty behaviour of the avatar. 
In this regard, frustration is the subscale with the highest variability. Due to the small sample, one or two individuals who decisively dislike the system already cause a strong distortion. Considering this fact, and the fact that mental demand was on average higher than frustration, we think that the overall workload is content-driven, not system-driven. 

\subsubsection{Exploratory Correlation Analysis}
For each participant, task accuracy and mean response time were computed across the 10 trials. Descriptive statistics are reported in Table~\ref{tab:performance_results}.

\begin{table}
\centering
\begin{tabular}{lcccc}
\hline
Measure & Mean & SD & Min & Max \\
\hline
Accuracy (0--10) & 4.00 & 1.20 & 2 & 6 \\
Mean Response Time (s) & 32.50 & 23.29 & 8.28 & 81.51 \\
\hline
\end{tabular}
\caption{\label{tab:performance_results}Descriptive statistics for task performance across 10 trials.}
\end{table}

Participants achieved an average accuracy of 4 correct responses out of 10 trials. Mean response times showed considerable variability across individuals.
To explore potential associations between perceived workload and objective performance, Spearman’s rank-order correlations were computed between the overall TLX score and both accuracy and mean response time.
No statistically significant association was observed between TLX total score and accuracy ($\rho = -0.22$, $p = .42$). Similarly, the correlation between TLX total score and mean response time was not significant ($\rho = 0.18$, $p = .53$).
These results indicate that, within the present sample, perceived workload was not significantly associated with either task accuracy or response time.

\section{Discussion}\label{sec:discussion}
In this work, we propose a general design methodology to consider the process of designing AI-driven avatars in VR for archaeological applications as a composition of conceptual design, logical design and the technical implementation, where our design methodology is a subset of the overall development life cycles for software described in literatures~\cite{acharya2020software}. We put a special emphasis on the conceptual design, divided it into (1) application realm, (2) avatar personality, and (3) information input/output modalities under the assumption of a VR-based deployment. We provide a comprehensive overview of various design considerations with our conceptual requirement space model to help system designers to navigate the vast design space that originates in complex setups such as RAG-driven VR avatars for archaeological purposes. Furthermore, we created a prototype while considering that overview and conducted an evaluation of the prototype with a number of different metrics.

The specific set of requirements we chose is a presentation (teaching)-focused system for an individual user who is either a student or a lay experienced user.
The avatar's personality was chosen to be either an expert or a semi-expert, who has an authorial narration style and is visualised abstractly as a more abstract, robotic human-like figure.
By blending abstract-robotic and human-like, it becomes apparent that the distinction we make between abstract and human-like in our presented requirement space is fluid and interpretative.
The proposed classification could therefore be understood as a heuristic framework rather than a rigid taxonomy.

As for the prototype, we implemented the interaction to be audio-focused, namely through STT for input and TTS for output. Once the user utters a question, this question gets forwarded to a RAG pipeline for answer generation. Once the query is done, the answer is returned from the RAG to the avatar in VR and relayed from the avatar to the user. While the avatar speaks, it shows a talking animation. The historical/archaeological topic the prototype addresses is a historically grounded VR reconstruction of the mausoleum of Maxentius at the Via Appia Antica in Rome. Through the integration of avatars in such archaeological environments, they become an interface to acquire knowledge about digitised cultural heritage, while being immersed in environments that use that cultural heritage.
This helps users to not simply acquire AI-driven knowledge through such avatars, but they can simultaneously see the visual evidence that belongs to the information from literature, helping them to interpret what they see and integrate theoretical knowledge into the mental image they have about a monument.

Furthermore, through the combination of the user's presence in an immersive environment and interaction through speech, we designed the prototype in a way that increases engagement with the system and the topic to make use of heightened presence for increased information retention.
And this increased engagement is what our user study suggests while users interact with the avatar. The workload with the avatar is perceived to be moderate to low, and the generally higher mental demand paired with the average frustration level indicates a healthy, task-driven engagement in general.

In addition, the metrical answer quality assessment of the system demonstrates that metadata-based query manipulation, which is helpful for future system designs to allow system steering through non-technical domain experts, improves answer quality when combined with sufficiently large embedding and LLM models. While BERTScore indicates high semantic similarity (77–79\%), qualitative evaluation via a LLM-as-judge revealed limited answer depth compared to expert ground truth responses.

Those outcomes are already very promising, despite the limitations of our work. For one, the classification into application realm, personality and I/O of the avatar is a primarily literature-synthesised system and remains empirically unvalidated across multiple use cases. Also, some of the characterisations for the system, such as human charactertypes, are from humanities research. Trying to directly translate real-human character traits to a digital system is in danger of oversimplifying a complex problem, which again points to the fact that the current conceptual requirement space is only to be understood as a heuristic framework.

Also, the QA pairs were derived from a small expert sample and encompass only 10 pairs of questions and answers. While this is enough for a prototype, a distribution-level system requires much more data to be properly evaluated. Furthermore, the selection of the underlying data, as well as the relevance scale for metadata steering introduce human bias into the retrieval, which could be considered a limitation. However, we consider our approach to be more general in the sense that, for example, exhibitions sometimes want to present a specific narrative perspective for which a biased system is required. For the example of the Maxentius mausoleum, we want the answers to focus highly on the mausoleum itself. Thus, introducing a bias that steers the retrieval into that direction is by design.

A further limitation concerns the treatment of visual uncertainty and reconstruction authenticity. 
While the system delivers the possibility to retrieve scholarly knowledge, it does not yet explicitly communicate which aspects of the virtual reconstruction are hypothetical or lack evidence. 
In digital archaeology, visual reconstructions inevitably embed interpretative decisions and the current prototype does not systematically encode or visualise degrees of evidential certainty, nor does it reference competing archaeological hypotheses to alternative visual interpretations of the monument. 
As a result, the status of visual elements could remain hidden rather than transparent to the user.

Lastly, our evaluation is a composition of different metrics. The problem is that our very specific use case has no one standardised evaluation system. Therefore, we used the semantic BERTscore assessment as a  complement to the LLM-as-a-judge approach for answer depth assessment, and a user study for the usability part. 

Building upon those limitations, our future research may take a number of different directions. 
First of all, we like to integrate a way to present the uncertainties of made statements to users. We want to integrate enhanced gestures into our avatar to increase the social presence of users and, possibly, even integrate adaptive personality systems that adjust themselves, depending on the user's expertise.
Also, future work should integrate more visual highlighting, inside the 3D environment, of not only uncertainties, but also various layered visualisations, such as different reconstruction hypotheses. Such integrations would allow users of the system to really access dynamic immersive spaces with a low interaction threshold and a high information density. Lastly, the evaluation methodology could be expanded and possibly even generalised. This would help in creating a more generalised system to expand on further monuments, and to create no-code interfaces to help curators and policy makers to deploy them.

\section{Conclusions}\label{sec:conclusions}
With this work, we created a RAG-driven VR avatar prototype for the Maxentius mausoleum at the Via Appia Antica in Rome as a use case, where the avatar is an interface for users to have access to expert-level knowledge about the monument. Through this, users find themselves situated in an immersive visual reconstruction of the monument and have access to a knowledge base simply by asking questions verbally, helping them to create a connection between theoretical facts and visual evidence.

We, furthermore, present a first concise elaboration of the various design considerations of such AI-driven archaeological avatars in immersive environments. We describe the three main building blocks to consider for such avatar systems, which are the general systems’ scope, the personality of the avatar and the way users can interact with them.

Also, we conducted an evaluation of our prototype's performance in terms of answer quality and user workload to identify strengths and weaknesses that may result from the use case. To do that, we calculated different answer performance metrics in different setups, and we posed Maxentius mausoleum-related expert-level questions to non-experts in a NASA-TLX workload study to see how difficult users perceive such systems. We found out that users perceive such systems as engaging and the workload as manageable, albeit their performance in answering tasks does not directly reflect on their perceived success. 

Given the complexity of the topic and the current stage of the prototype, there is ample opportunity for further developments and adaptations. However, with this work, we took an important step toward systematically transitioning a very promising synthesis of different technologies to the field of virtual archaeology.

\begin{acks}
We acknowledge funding by DFG, project ID 251654672–TRR 161 and by Ce.Di.Pa. - PNC Programma unitario di interventi per le aree del terremoto del 2009-2016 - Linea di intervento 1 sub-misura B4 - ``Centri di ricerca per l'innovazione'' CUP J37G22000140001. 
\end{acks}

\bibliographystyle{ACM-Reference-Format}
\bibliography{bib}

\end{document}